# Field Inversion Symbolic Regression with Embedded Equation Learner for Interpretable Turbulence Model Correction

Li Jiazhe[a], Wu Chenyu[a], He Zizhou[a], Zhang Yufei[a, b, c, ]*

[a] School of Aerospace Engineering, Tsinghua University, Beijing, 100084, China

[b] State Key Laboratory of Advanced Space Propulsion, Tsinghua University, 100084, Beijing

[c] Caofeidian Laboratory, 063200 Tangshan, China

* Correspondence: zhangyufei@tsinghua.edu.cn

**Abstract**: An interpretable, physics-consistent turbulence model correction framework, termed FISR-Equation Learner (EQL), is proposed by embedding equation learning directly into a Partial Differential Equations (PDE)-constrained field inversion process based on the adjoint method. Unlike conventional two-stage approaches, the correction model is optimized end-to-end in parameter space using an EQL architecture, enabling the direct identification of compact analytical expressions while maintaining consistency with the governing equations. The method is applied to the shear-stress-transport (SST) model and trained on two canonical separated flows, the curved backward-facing step and the NASA hump. The resulting explicit expression significantly reduces separation bubble overprediction and improves reattachment prediction, achieving performance comparable to neural-network-based end-to-end methods while retaining full interpretability. Generalization is demonstrated on unseen configurations, including periodic hills, a surface-mounted cube, and the high-lift NLR7301 airfoil. The model improves separated-flow predictions and stall characteristics without degrading attached boundary-layer performance. Overall, FISR-EQL provides a practical pathway toward optimal yet transparent data-driven turbulence model correction.

**Keywords**: Data driven, Turbulence modeling, Field inversion, Interpretable machine learning, Online Learning, Equation Learner

## 1. Introduction

The development of reliable turbulence models has been and remains a key underlying requirement for aerodynamic design and analysis of flight vehicles. Simulations based on Reynolds-averaged Navier–Stokes simulations are commonly performed for the prediction of aerodynamic force, moment, and flow separation behavior over a wide range of operating conditions. The accuracy of such simulations is known to significantly affect important flight designs in terms of performance, stability, control, and loading behavior of flight transport designs. However, despite their computational efficiency, traditional turbulence models based on Reynolds-averaged Navier–Stokes (RANS) simulations exhibit restrictive behavior when simulating typical flow patterns involved in flight designs.

This limitation arises from the simplified physical assumptions embedded in the underlying structure of RANS models, which lead to "model inadequacy"; errors in the underlying structure of the model may result in inadequacies even if the best possible set of parameters has been inferred.[1] For example, it is well established that RANS turbulence models often perform poorly in representing separation bubble physics, especially downstream reattachment and recovery.[2] RANS models often underestimate Reynolds stress in the separated shear layer, which is crucial for accurately predicting flow separation.[3, 4] This underestimation results in insufficient viscous effects in the shear layer, causing classic turbulence models, such as the shear-stress-transport (SST) model, to consistently overestimate the size of separation bubbles. These deficiencies highlight a fundamental limitation of traditional RANS modeling approaches: improving model performance through parameter calibration alone is often insufficient when the dominant source of error stems from the model's structural assumptions rather than uncertainty in its coefficients. As a result, there is growing interest in augmenting or correcting RANS models using external information that can compensate for missing or misrepresented physics, particularly in complex flow regimes such as separation.

Using experimental data or direct numerical simulation (DNS) data to correct turbulence models, namely data-driven approaches, has become increasingly popular in recent years. Many representative studies have advanced data-driven turbulence modeling, including uncertainty quantification[5], modeling and analysis of Reynolds stress tensor prediction errors[6-8], and more. The field inversion and machine learning (FIML) framework[9,10] proposed by Duraisamy's team provides a novel approach for the optimization of turbulence models. The distinctiveness of the FIML approach lies in its objective to establish machine learning (ML) models that correlate local flow characteristics with model correction factors, which are then employed to refine turbulence models for new flow conditions.[11] By introducing correction factors into standard turbulence model equations and leveraging these factors to modify predictive performance, this method achieves model consistency. Using FIML methods, substantial research has been conducted to improve turbulence model performance.[12-18] Building on this, many studies have sought to address the inherent limitations of FIML, particularly in terms of generalizability, nonlocality, and interpretability. For example, Bin et al.[19] proposed a constrained recalibration framework that restricts model corrections to degrees of freedom that do not affect fundamental physical calibrations, thereby addressing the generalization limitations of data-driven approaches such as FIML. Wu et al.[20] extended the FIML framework by introducing a transport-equation-based nonlocal model, thereby overcoming the limitation that corrections depend only on local flow features.

In addition, to address the black-box nature of neural-network-based FIML models, Wu et al. proposed the Field Inversion and Symbolic Regression (FISR) framework[21,22], which replaces the neural-network-based ML stage in FIML with symbolic regression, thereby overcoming the black-box limitations of neural networks. Symbolic expressions offer significant advantages in terms of interpretability and generalizability compared with neural-network models. He et al.[23, 24] applied FISR as well as field inversion and physics-driven-classification modeling (FI-PDCM) to enhance the prediction accuracy of the Spalart–Allmaras (SA) model for airfoil stall, indicating that the new model improves predictive capability for airfoil stall without compromising the performance of the baseline SA model for attached flows.

The ultimate goal of turbulence model correction is to obtain a modification that is model-based and case-independent, i.e., a correction that generalizes across different flow configurations. Conventional FIML and FISR frameworks decompose this objective into two stages: field inversion (FI) followed by machine-learning-based model regression (ML) or symbolic regression (SR). Although this approach has proven effective in many problems, it also has certain limitations. This decomposition implicitly introduces two separate optimization problems: first, identifying a spatially distributed correction field that makes the RANS solution best match reference data; second, fitting a functional model that minimizes the discrepancy between its pointwise predictions and the inferred correction field. The resulting two-stage procedure accumulates errors from both steps and, crucially, does not yield the model that directly minimizes the RANS mismatch with data in the space of model parameters, which is referred to as objective mismatch. Moreover, models obtained through such a decoupled, two-stage training paradigm sometimes exhibit numerical instability when coupled back into the Computational Fluid Dynamics (CFD) solver, leading to divergence or failure to achieve the expected correction effect, thereby undermining their practical applicability.

To overcome these limitations, Holland et al.[25,26] introduced the FIML-direct framework, in which a neural network is trained as the model parameter within the inversion process. Although the strong nonlinearity of neural networks poses additional challenges for optimization, this end-to-end formulation improves predictive performance. More importantly, this embedded-model formulation is indispensable for more complex unsteady systems, since in unsteady problems it is difficult to define a time-dependent intermediate correction field as a surrogate; consequently, the correction model must be optimized directly. However, the FIML-direct approach relies on neural networks to represent the correction model, which, as discussed earlier, is inherently less interpretable and may exhibit weaker generalization capability compared with symbolic regression–based formulations. Jäckel[27] combined FIML-classic and FIML-direct by first using the FIML-classic approach to identify regions that are most critical to the target correction and training a Radial Basis Function (RBF) model in those regions. Subsequently, the FIML-direct approach was applied for further tuning across multiple cases. While the RBF model possesses a certain degree of interpretability and stability, it is still less intuitive than explicit expressions.

The present study combines the complementary strengths of these two paradigms: the parameter-space optimality of FIML-direct and the interpretability afforded by two-stage approaches. We propose the FISR-EQL framework, in which the Equation Learner (EQL) architecture is embedded directly into the FIML-direct process, replacing conventional neural networks with an expression-oriented EQL network. EQL[28-30] is a neural-network-based symbolic architecture, as illustrated in Figure 1(a), in which conventional activation functions are replaced by symbolic operators commonly used in SR. These include unary operators (e.g., sin, cos) and binary operators (e.g., multiplication), enabling the network to represent explicit analytical expressions. To obtain a compact and interpretable formulation, sparsity constraints are imposed on the network parameters during optimization; the specific implementation is described in Section 2.3. Once the network becomes sparse, the analytical expression represented by the model can be directly extracted from the small set of remaining parameters, as illustrated in Figure 1(b). This enables end-to-end identification of explicit analytical corrections, representing a fundamental advance in interpretable turbulence model correction.

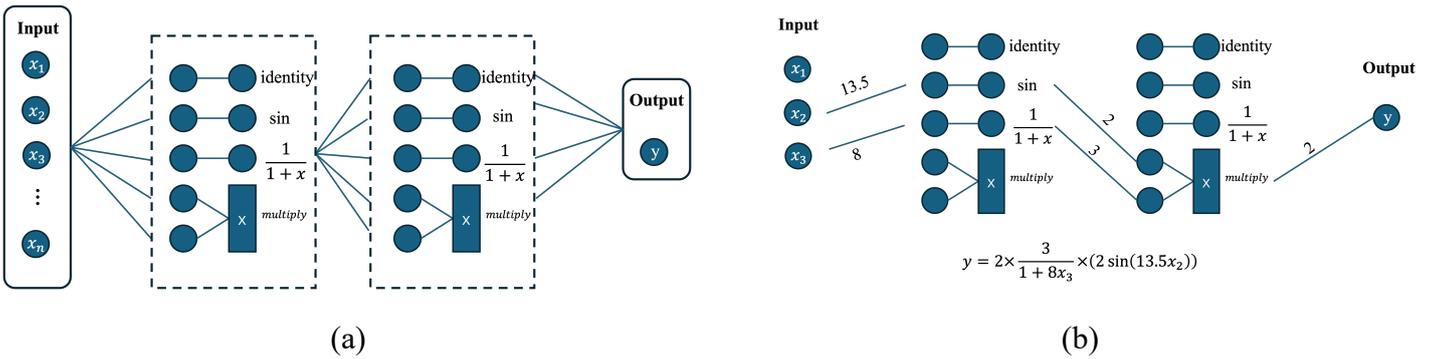

Figure 1 EQL Framework

Table 1 shows the differences among these FIML methods and the advantage of the proposed FISR-EQL framework. Among these four approaches, only the proposed FISR-EQL framework is capable of producing an explicit analytical expression without incurring objective mismatch. Compared with other end-to-end online approaches, such as Kalman-filter-based parameter updates[31,32] or gene expression programming[33], the proposed method is able to leverage discrete adjoint techniques to perform gradient-based optimization efficiently, yielding substantially improved computational efficiency.

Table 1 Comparison of different FIML methods

| Methods | Description | Objective Mismatch | Obtained Model |
|---|---|---|---|
| FIML-classic[9,10] | A two-stage framework in which a spatial correction field is first inferred via field inversion and subsequently regressed offline into an explicit model using machine learning. | Yes | Neural Networks |
| FIML-direct[25,26] | An end-to-end formulation in which a neural network–parameterized correction model is embedded within the RANS equations and optimized directly under PDE constraints. | No | Neural Networks |
| Two-stage FISR[21-23] | A two-stage approach in which symbolic regression is applied offline to fit an explicit analytical expression to a previously inferred correction field. | Yes | Explicit expression |
| FISR-EQL （present work） | An end-to-end strategy in which a symbolic-regression-based model is embedded directly into the governing equations and optimized in a single PDE-constrained framework without introducing an intermediate correction field. | No | Explicit expression |

# 2. Methodology

In the present work, this approach is applied to the SST turbulence model to improve the performance of separation flow predictions. In this chapter, the baseline turbulence model and the FISR-EQL method are discussed.

**2.1 Turbulence Model**

Our baseline turbulence model is Menter's $k-\omega$ SST model[34], proposed in 2003. The corresponding formulations are not repeated here for brevity. In the present work, the correction is introduced by multiplying the production term $P_k$ by a factor $\beta$:

$$\frac{\partial(\rho k)}{\partial t} + \frac{\partial(\rho U_i k)}{\partial x_i} = \beta P_k - \beta^* \rho k \omega + \frac{\partial}{\partial x_i}\left[(\mu + \sigma_k \mu_t)\frac{\partial k}{\partial x_i}\right] \quad (1)$$

If $\beta \equiv 1$ is applied throughout the entire flow field, the model performs exactly the same as the original SST model. If $\beta > 1$ in some region, the value of $k$ increases because $P_k > 0$, and therefore $\nu_t$ increases. In this way, the flow field can be controlled by modifying the value of $\beta$ in different regions, as will be shown in the following sections. In conventional two-stage FISR and FIML-classic frameworks, the correction factor $\beta$ is treated as an explicit function of the spatial coordinate $x$: the first stage identifies the spatial distribution of modeling error, and the second stage constructs a model for this distribution. In contrast, in FIML-direct and in the FISR-EQL framework developed in this work, $\beta$ is formulated as a function of local flow features $q$ and is evaluated directly from the instantaneous flow field.

The SST model predicts zero-pressure-gradient and attached flat-plate boundary layers with high accuracy. Therefore, the correction should not be activated within attached boundary layers, as this would deteriorate predictions for boundary-layer-dominated quantities such as skin-friction drag. Following Ref.22, a shielding function is applied to the original $\beta_{raw}$:

$$f_d = 1 - \tanh[(8r_d)^3], \quad r_d = \frac{\mu + \mu_t}{\rho \kappa^2 d^2 \sqrt{u_{i,j} u_{i,j}}} \quad (2)$$

This shielding function is typically zero inside the boundary layer and unity elsewhere. In FISR-EQL and FIML-direct, $\beta_{raw}$ denotes the direct output of the EQL and neural-network models, respectively. In all cases, the $\beta$ appearing in Eq. (1) is obtained by applying the following transformation to $\beta_{raw}$ before it is introduced into the production term:

$$\beta = (\beta_{raw} - 1)f_d + 1 \quad (3)$$

so that the boundary-layer solution remains unaffected regardless of the raw output of $\beta$. Consistent with the end-to-end philosophy, the shielding function is incorporated directly during training, ensuring consistency between training and deployment. This treatment is also adopted in Ref.22, where it is referred to as conditional FI.

## 2.2 Conventional FIML Framework

Before introducing the EQL formulation, we first review the conventional FIML framework. For FIML-classic and two-stage FISR, a minimization problem is solved:

$$\min_{\boldsymbol{\beta}} J(\boldsymbol{\beta}, \boldsymbol{w}) = \min_{\boldsymbol{\beta}}\left[\lambda_{QoI}\sum_{i=1}^{K}(v_i - \hat{v}_i(\boldsymbol{\beta}))^2 + \lambda_{L2}\sum_{j=1}^{N}(\boldsymbol{\beta}_j - 1)^2\right] \text{ subject to } \boldsymbol{R}(\boldsymbol{\beta}, \boldsymbol{w}) = 0 \quad (4)$$

$v_i$ refers to the experimental value or DNS and Large Eddy Simulation (LES) result of flow-field variables, such as velocity and pressure and $\hat{v}_i(\boldsymbol{\beta})$ refers to the RANS simulation result, which is a functional of the beta distribution function $\boldsymbol{\beta}(x)$. Therefore $\sum_{i=1}^{K}(v_i - \hat{v}_i(\boldsymbol{\beta}))^2$ represents the quantity of interest (QoI). $R(\boldsymbol{\beta}, \boldsymbol{w}) = 0$ means that the $\boldsymbol{\beta}$ and $\boldsymbol{w}$ distributions must satisfy the governing equations, namely the Navier–Stokes equations including turbulence equations. The objective is to minimize the discrepancy between experimental values and RANS simulation results. Here, $K$ data points are used, indicating that data are not required at all grid points but only at sparse locations. The other term in the objective function is a regularization term that reduces the deviation of $\boldsymbol{\beta}(x)$ from 1, since a simple $\boldsymbol{\beta}(x)$ distribution is desired. An overly complex $\boldsymbol{\beta}(x)$ distribution makes it difficult to model $\boldsymbol{\beta}$.

The adjoint method[35] is used to solve this minimization problem and is implemented using the open-source software DAFoam[36]. The optimization loop proceeds as follows:

1. Primal solve: Solve $R(\boldsymbol{\beta}, \boldsymbol{w}) = 0$ for $\boldsymbol{w}$ given $\boldsymbol{\beta}(x)$.
2. Objective evaluation: Compute $J(\boldsymbol{\beta}, \boldsymbol{w})$.
3. Adjoint solve: Solve $\left(\frac{\partial R}{\partial w}\right)^T \boldsymbol{\psi} = -\left(\frac{\partial J}{\partial w}\right)^T$
4. Gradient evaluation: Compute $\frac{dJ}{d\beta} = \frac{\partial J}{\partial \beta} + \boldsymbol{\psi}^T \frac{\partial R}{\partial \beta}$
5. Design update: Use an optimization algorithm (e.g., SLSQP) to update $\boldsymbol{\beta}(x)$ using the computed gradient.

This process is iterated until convergence of the objective function or satisfaction of optimality conditions.

**2.3 FISR-EQL Method**

Compared with the two-stage FISR strategy, which introduces additional error and potential instability due to its decoupled treatment, the proposed FISR-EQL framework involves only a single QoI, namely the flow-field mismatch, by embedding the EQL model directly within the inversion process. A single PDE-constrained optimization (comprising multiple iterative updates) yields the final mapping from local feature variables $q$ to the correction factor $\beta$. Under a prescribed EQL architecture and optimization algorithm, this model is optimal with respect to the chosen QoI. In contrast, the two-stage FISR approach does not guarantee optimality with respect to the original QoI due to the objective mismatch introduced by the two-stage procedure. While FIML and FISR ensure consistency at the turbulence-model level, the FISR-EQL framework further establishes consistency at the level of the correction model itself, as illustrated in Figure 2.

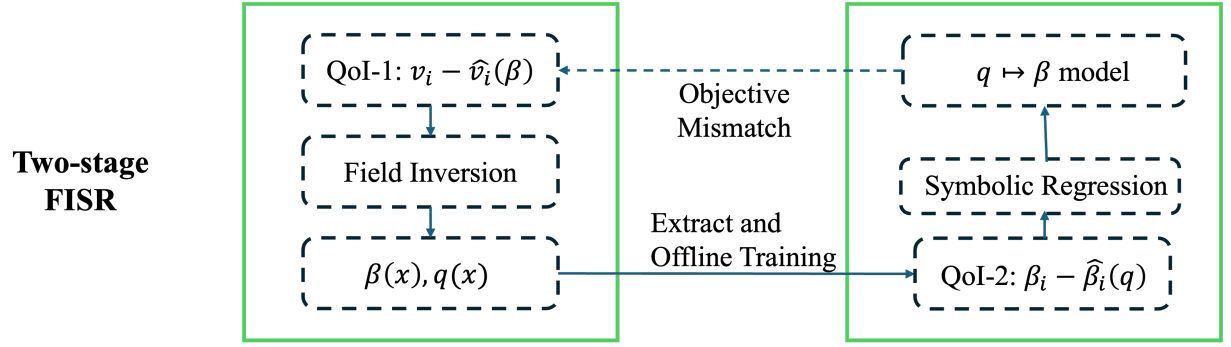

(a) Two-stage FISR framework

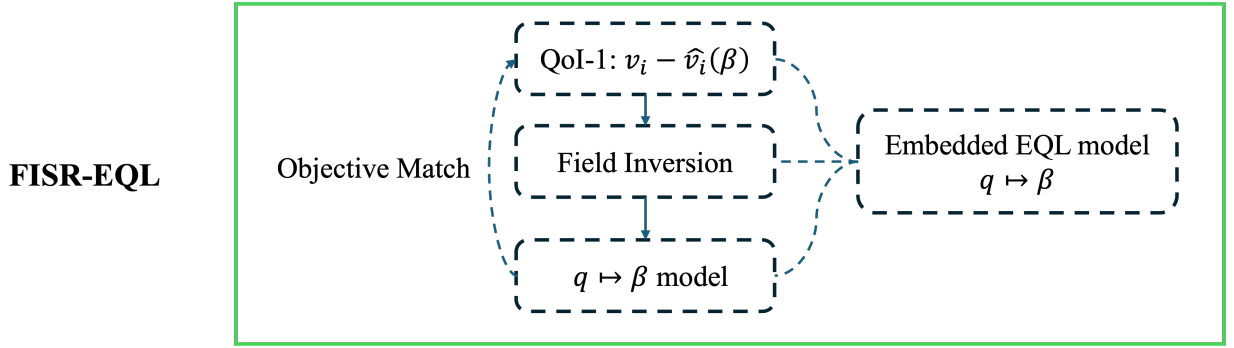

(b) FISR-EQL framework

Figure 2 Comparison of two FISR methods

Within the FISR-EQL framework, an EQL network is embedded as an integral component of the turbulence model and optimized during field inversion, thereby unifying the two-step process of error source identification and error modeling into a single optimization problem. In the present work, the structure of the EQL network and its role in the overall optimization procedure are illustrated in Figure 3.

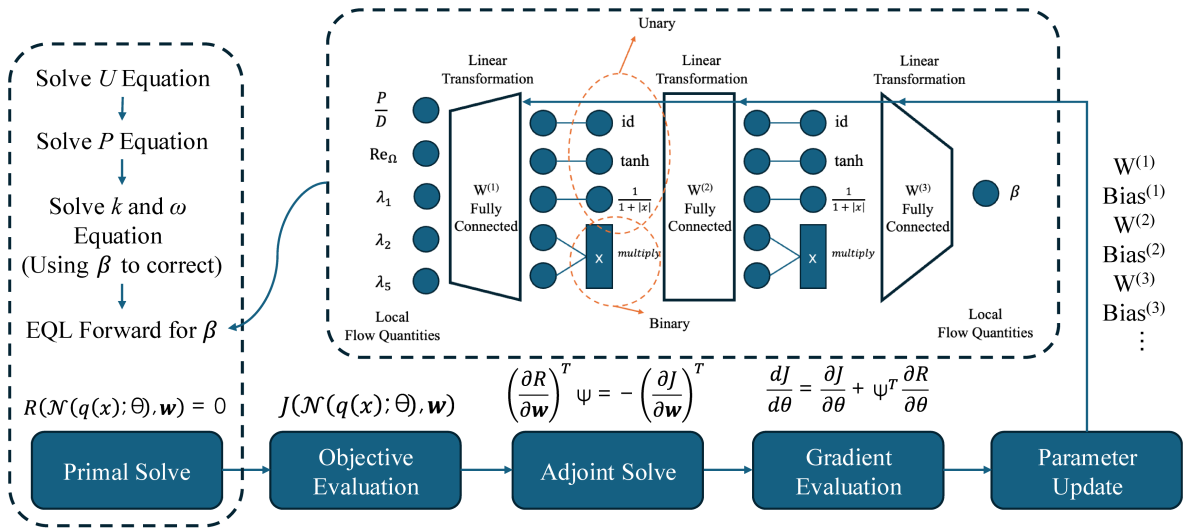

Figure 3 Structure of FISR-EQL framework

The EQL network takes local flow features $q$ as inputs, listed in Table 2. These features are selected based on the physical insights and modeling strategies discussed in Ref.22. They characterize the local state of the flow and are mapped by the EQL network to the required turbulence model correction variable $\beta$. Since the EQL output enters the turbulence closure, it directly affects the transport of turbulent quantities during CFD simulation. As a result, the mismatch between the RANS solution and the reference data (LES in the present study) becomes an implicit function of the EQL parameters, thereby transforming the problem into a single PDE-constrained optimization: the parameters of the EQL network are optimized to minimize the discrepancy between the predicted and reference flow fields.

Table 2 Feature Selection

|  |  |
|---|---|
| $\dfrac{P}{D}$ | This is the ratio of the production to dissipation of turbulent kinetic energy. As discussed in the introduction, this ratio is typically much greater than one in free shear layers, indicating possible overproduction of eddy viscosity. Thus, it may serve as an effective indicator for correction. |
| $Re_\Omega$ | Originally proposed in[40], this vortex-based Reynolds number is designed to identify free shear regions, providing spatial localization for model discrepancies. |
| $\lambda_1, \lambda_2$ and $\lambda_5$ | These represent the first, second and fifth invariants of the dimensionless strain-rate and rotation-rate tensors, respectively, as introduced by Pope[41]. They offer coordinate-independent characterization of local flow topology and have been widely used in turbulence modeling and flow feature extraction. |

From an architectural perspective, the EQL network shares the same feed-forward backbone as a conventional multilayer perceptron (MLP), consisting of multiple linear transformation layers parameterized by weight matrices $W$ and associated biases. The key distinction lies in the activation mechanism. Instead of employing a uniform nonlinearity such as Rectified Linear Unit (ReLU), each layer is composed of a heterogeneous set of unary and binary operators, in a manner analogous to SR. In the present work, three unary and one binary operators are used, as listed in Table 3.

Table 3 Operators used in EQL

| Type | Operator | Detail |
|---|---|---|
| Unary | $\text{id}(x)$ | Identity: Preserves the input |
| | $\tanh(x)$ | Hyperbolic tangent: Zero at the origin and saturates to $\pm 1$ for large-magnitude inputs. ($\tanh x = \frac{e^x - e^{-x}}{e^x + e^{-x}}$) |
| | $\frac{1}{1+|x|}$ | Evaluates to unity at zero and decays toward zero for large inputs. |
| Binary | $x \cdot y$ | Computes the product of two inputs. |
| | $x + y$ | Addition: implemented only within the linear transformation and not treated as a separate activation operator. |
| | $x - y$ | Subtraction: implemented only within the linear transformation and not treated as a separate activation operator. |

In the present EQL formulation, division is not introduced as an independent operator. On the one hand, incorporating division as an activation operator can lead to severe training instability, since the denominator may approach zero during optimization, an issue that is particularly pronounced when sparsification is enforced. On the other hand, the operator $\frac{1}{1+|x|}$ allows input variables to appear in the denominator in a numerically stable manner, enabling the construction of division-like expressions without explicit division. Although alternative strategies have been proposed, such as outputting the numerator and denominator separately in the final layer to realize rational functions[37], the focus of this work is to explore the application of EQL within the FISR framework. Therefore, a relatively simple and robust EQL architecture is adopted.

A central advantage of the proposed approach is that it ultimately yields an explicit analytical expression; consequently, the compactness and interpretability of the resulting model are of paramount importance. The compactness of the learned model can be examined through the coefficient matrix $W$ of the fully connected layers shown in Figure 3. In a conventional neural network, $W$ is an unconstrained dense matrix. However, to obtain a concise analytical expression, it must evolve into a sparse matrix with only a few nonzero entries. Such sparsity can be promoted by imposing regularization constraints on both $W$ and the bias terms during optimization. The proposed FISR-EQL framework follows the original EQL methodology and achieves this objective through a three-phase training procedure. During this process, the heat map of the coefficient matrix $W$ evolves from a dense pattern to a sparse structure, ultimately fulfilling the goal of SR, as illustrated in Figure 4.

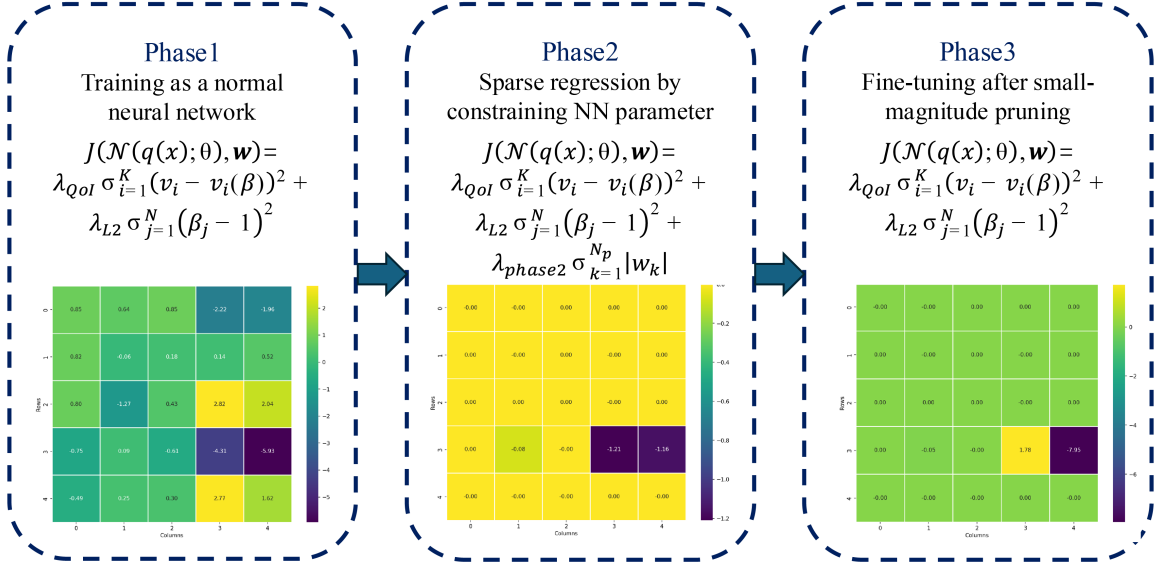

Figure 4 Training procedure of EQL framework

In the first stage, the network is trained in a conventional manner, with an objective function that includes only the data misfit at observation locations and an $L_2$ regularization term on the inferred $\beta$ field. In the second stage, an $L_1$ penalty on the network parameters is introduced, enforcing sparsity while preserving predictive performance. This drives the network toward a parsimonious representation by suppressing weakly contributing connections. In the third stage, the parameter regularization is removed. Based on the parameter distribution obtained in Stage 2, a threshold is selected and applied to prune branches whose contributions to $\beta$ are negligible. In practice, this is implemented by setting parameters below the threshold to zero, which effectively removes the corresponding branches from both the forward and backward passes without affecting the remaining computations. The reduced network is then fine-tuned using the same objective as in Stage 1, with the sole purpose of reoptimizing the retained parameters. Through this multiphase procedure, the EQL network converges to a highly compact structure, from which a concise and interpretable analytical correction model can be extracted. Similar to the FIML-direct approach, the FISR-EQL framework treats the network parameters as the optimization design variables. Therefore, in all three training phases, the gradients of the objective function with respect to the network parameters are obtained using the adjoint-based automatic differentiation capability implemented in DAFoam[36].

# 3. FISR-EQL Model Training

## 3.1 Case Setup

Two canonical separated shear-flow configurations are employed to train the correction model in this study: the Curved Backward-Facing Step (CBFS)[38] and the wall-mounted hump (HUMP)[39]. Leveraging the DAFoam framework[36], these two cases are optimized in a unified manner. Specifically, at each optimization iteration, both cases are evaluated using the same set of EQL parameters. The objective functions for the two flows are computed independently, and their gradients with respect to the EQL parameters are obtained via the discrete adjoint method. The resulting gradients are then combined, with prescribed weights, to form the total gradient of the aggregated objective with respect to the EQL parameters. This multi-case training strategy enforces consistency across distinct flow configurations and promotes the identification of a correction model with improved generality. For both cases, the incompressible solver (DASimpleFoam) of DAFoam is utilized.

For the CBFS case, the geometric configuration and boundary conditions are illustrated in Figure 5. The Reynolds number, based on the step height $h = 1$ m, is set to $Re_h = 13,700$. The centerline inlet velocity is prescribed as about $1$ m/s. The streamwise velocity component is selected as the flow variable for assessing the effectiveness of the correction. In the separated region, the LES velocity field is sampled at 30 monitoring locations, which serve as reference data in the objective function.[42]

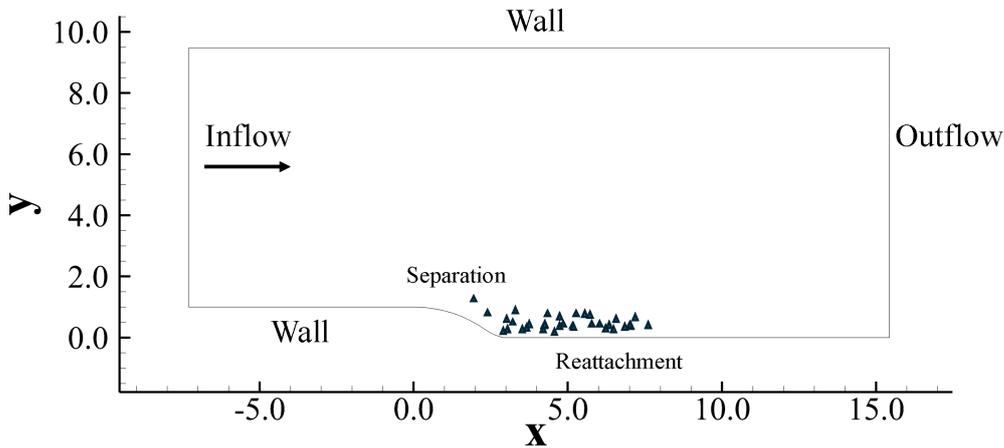

Figure 5 Geometry of CBFS case

For the NASA hump case, the geometry and sampling strategy are illustrated in Figure 6. The Reynolds number based on the chord length $c = 1$ m is $Re = 0.936 \times 10^6$, with an inlet velocity of $14.88 \, m/s$. The streamwise velocity component is also selected, with 40 monitoring locations in the hump case.[43]

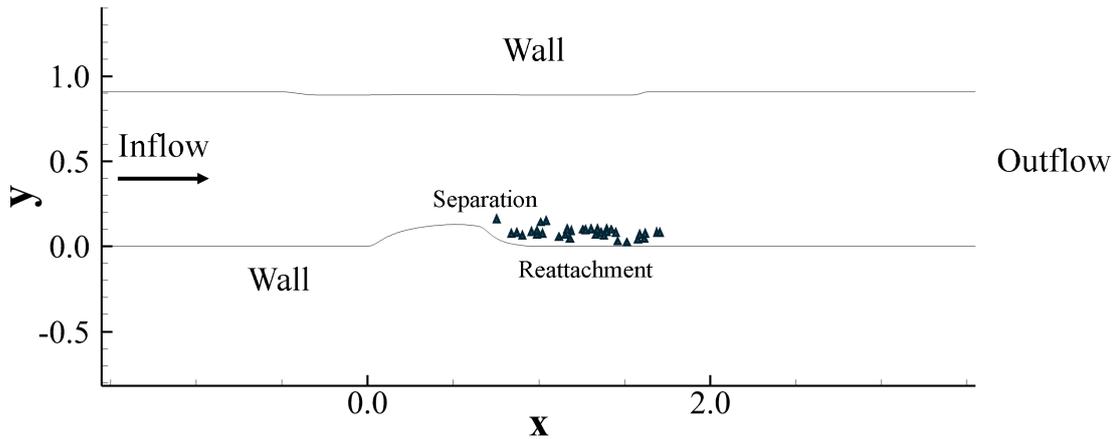

Figure 6 Geometry of Hump case

## 3.2 EQL Training

Three kinds of unary operators and one binary operator are used, as listed in Table 3. To assess the impact of the EQL network size on the training process and to eliminate potential biases introduced by a specific architecture, three networks of increasing complexity are considered. Each network is evaluated using 14 runs with random initialization, and the results are summarized in Table 4.

The Small network consists of a single EQL layer, including three unary operators (each of the three operators listed in Table 3 appears once) and two binary operators (multiplication), yielding three valid results (No. 1, 2, and 3).

The Medium network contains two EQL layers, each with three unary operators and two binary operators, resulting in three valid outcomes (No. 4, 5, and 6).

The Large network also consists of two EQL layers, but each layer includes six unary operators (each operator repeated twice) and three binary operators, producing three valid results (No. 7, 8, and 9).

We outline several principles that should guide the selection of expressions:

1. Interpretability. Since the primary purpose of using FISR-EQL is to obtain interpretable expressions, the source of the correction should be directly analyzable from the expression itself. For example, from the second and third expressions it can be seen that when the magnitude of negative $\lambda_2$ or positive $Re_\Omega$ increases, it generally indicates enhanced shear and therefore requires correction. In contrast, overly complex expressions such as Expressions 8 and 9 do not possess this level of simplicity.
2. Boundedness. The correction should avoid divergence in nearly all simulations. This requires that the correction should not become unbounded or produce values smaller than 1 where it should not, since this would reduce the eddy viscosity and may lead to convergence difficulties. For example, from Expression 6 it can be directly observed that when $\lambda_2$ approaches negative infinity, both fractions approach 0, and therefore the maximum value is 1.83. When $\lambda_2 = 0$, the expression reaches its minimum, which is close to 1.

3. Accuracy. In regions where no correction should be applied (for example, in uniform flow), the expression must not produce values different from 1; otherwise, it would affect flows that should be correctly computed without correction. In this study, this property is ensured through $L_2$ regularization on $\beta$. In addition, corrections should not be applied within the boundary layer, which is enforced by the shielding function $f_d$.

Under the premise that the above principles are satisfied, the expression with the smallest error is selected, based on the computed relative error of the reattachment location:

$$\varepsilon_r = \frac{\left|x_r^{\text{EQL}} - x_r^{\text{LES}}\right|}{\left|x_r^{\text{SST}} - x_r^{\text{LES}}\right|}, \varepsilon_{r,result} = \frac{\varepsilon_{r,cbfs} + \varepsilon_{r,hump}}{2} \quad (5)$$

we select the expression obtained from the fifth successful run for further analysis.

Table 4 EQL network structures and results

| No. | Output Equation | Relative Error |
|---|---|---|
| 1 | $\beta = 5.51 \times 10^{-2}\lambda_2 - 2.44 \times 10^{-2}\lambda_5 + 1.00$ | 17.20% |
| 2 | $\beta = -5.33 \times 10^{-2}\lambda_2 + 0.92$ | 28.71% |
| 3 | $\beta = (4.88 \times 10^{-6}Re_\Omega + 1.18)$ | 35.34% |
| 4 | $\beta = -0.615\tanh(1.56 \times 10^{-5}Re_\Omega + 0.449\lambda_2 + 3.77 \times 10^{-2}\lambda_5 + 5.28) + 1.61$ | 31.72% |
| 5 | $\boldsymbol{\beta = \left(\frac{-1.63}{1 - 2.10 \times 10^{-2}\lambda_2 - 1.92 \times 10^{-2}\lambda_5} + \frac{-1.08}{1 + |0.264\lambda_2 + 2.11 \times 10^{-2}\lambda_5 + 2.20|} + 2.95\right)}$ | **16.80%** |
| 6 | $\beta = \frac{-0.424}{1 + |9.82 \times 10^{-2}\lambda_2|} - \frac{0.458}{1 + |0.101\lambda_2|} + 1.83$ | 29.24% |
| 7 | $\beta = -\frac{1.47}{1 + |3.76 \times 10^{-2}\lambda_2|} - \frac{0.234}{1 + \left|5.19\frac{P}{D}\right|} + 2.61$ | 17.94% |
| 8 | $\beta = \left(\frac{0.820}{1 + |1.12 \times 10^{-2}\lambda_2|} + \frac{0.850}{1 + |6.35 \times 10^{-3}\lambda_2|}\right) \times \left(3.39 - \frac{1.72}{1 + |1.12 \times 10^{-2}\lambda_2|} - \frac{2.14}{1 + |6.35 \times 10^{-3}\lambda_2|}\right) + \frac{3.67 \times 10^{-2}}{1 + |1.12 \times 10^{-2}\lambda_2|} + \frac{0.173}{1 + |6.35 \times 10^{-3}\lambda_2|} + 1.50$ | 29.18% |
| 9 | $\beta = \left(0.342 + \frac{0.123}{1 + |2.24 \times 10^{-2}\lambda_2 - 7.67 \times 10^{-3}\lambda_5|}\right) \times \left(-2.83 \times \tanh(0.376\lambda_2 + 4.13) + 2.59 - \frac{0.765}{1 + |2.24 \times 10^{-2}\lambda_2 - 7.67 \times 10^{-3}\lambda_5|}\right) + 1.45$ | 19.36% |

Except for the EQL layers, an additional linear output layer is appended to produce the correction variable $\beta$. All training phases are optimized using the SNOPT algorithm[44], without early stopping. Figure 7 presents the objective-function convergence histories for all 42 training runs across the three phases. The selected fifth successful run is highlighted in red. The number of optimization iterations varies among runs and across phases; for runs that terminate a phase earlier, dashed lines are used to connect to the subsequent phase. A noticeable jump in the objective value occurs at the transition from Phase 1 to Phase 2 due to the introduction of the $L_1$ regularization on the network parameters. Conversely, a sudden decrease is observed when transitioning from Phase 2 to Phase 3, as the regularization term is removed. Compared with other runs, the selected case exhibits relatively small objective perturbations during the Phase 1–Phase 2 transition, indicating that the network already attained a relatively sparse structure after Phase 1. Consequently, the subsequent sparsification in Phase 2 proceeds more smoothly and efficiently.

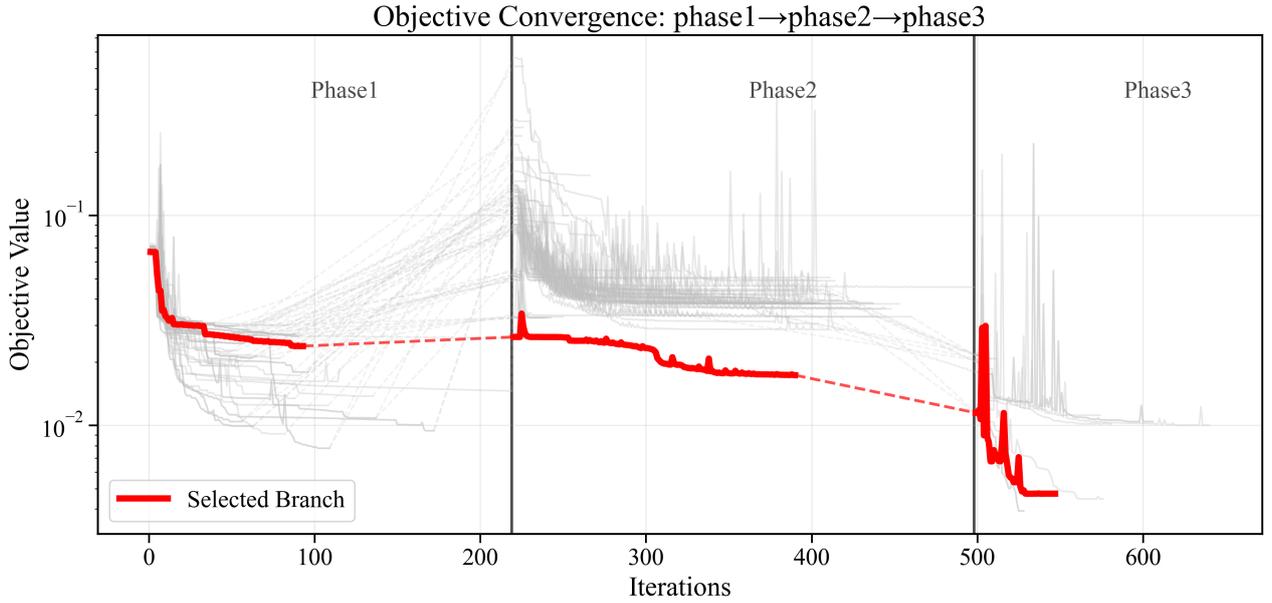

Figure 7 Objective Function Convergence (All runs plotted)

The objective function in this study comprises multiple competing terms: the streamwise velocity mismatch (QoI) for the CBFS case, the streamwise velocity mismatch for the HUMP case, the $L_2$ regularization of the $\beta$ field in both cases, and, during Phase 2, the $L_1$ regularization of the network parameters, resulting in a total of five contributions. The relative influence of these terms is controlled through weighting coefficients, which require careful tuning. Inappropriate weighting may lead to undesirable outcomes, such as: (1) an excessively strong $L_2$ regularization on $\beta$, which drives the EQL network to collapse to a constant output of unity, resulting in negligible reduction of the QoI and effectively no model correction; (2) an overly strong $L_1$ regularization on the network parameters, which forces the EQL network to produce a constant value significantly larger than unity. In this case, the QoI reduction is achieved through an extremely sparse structure, where only the bias in the final layer remains nonzero, rendering the $L_2$ regularization on $\beta$ ineffective. After extensive preliminary tuning, the weighting coefficients were ultimately set to $\lambda_{\text{CBFS}}^{\text{QoI}} = 10.0$, $\lambda_{\text{HUMP}}^{\text{QoI}} = 0.014$, $\lambda_{\beta}^{L_2} =$ $0.5, and\ \lambda^{L_1}_{\text{network}} = 0.005$. Using this weighting set, 9 cases listed in Table 4 successfully output equations in all 42 runs.

### 3.3 Performance on Training Cases

We evaluate the selected expression on the training configurations, CBFS and HUMP. The primary metric is the reattachment location. In addition to the baseline SST model and the proposed FISR-EQL approach, three reference strategies are also examined: pure FI, the FIML-classic method that applies ML to inversion data, and the FIML-direct method that optimizes an embedded neural network. For both FIML-direct and FIML-classic, the machine-learning component is implemented using a four-layer MLP with a "5(input)–20–40–20–1(output)" architecture. The objective function is defined identically to that used in Phase 1 of FISR-EQL, and optimization is carried out using the SNOPT algorithm.

The resulting skin-friction coefficient $C_f$ distributions for both training cases are shown in Figure 8. Figure 9 presents the streamwise velocity profiles at several downstream locations in the separated region. It can be observed that the corrected model brings the RANS predictions into closer agreement with the LES results, demonstrating the effectiveness of the proposed modification.

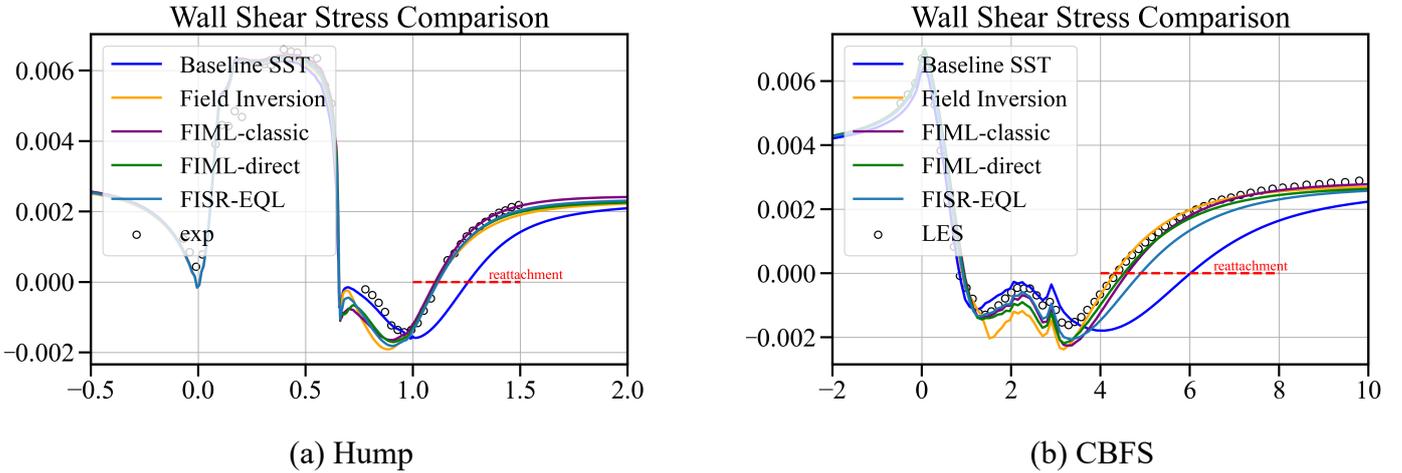

(a) Hump  (b) CBFS

Figure 8  $C_f$ distribution of two training cases

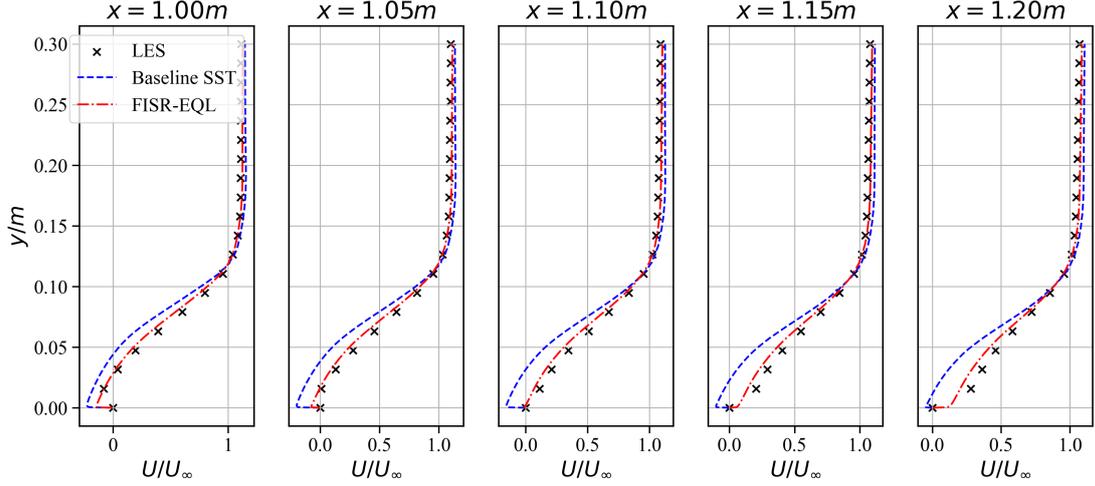

(a) Hump

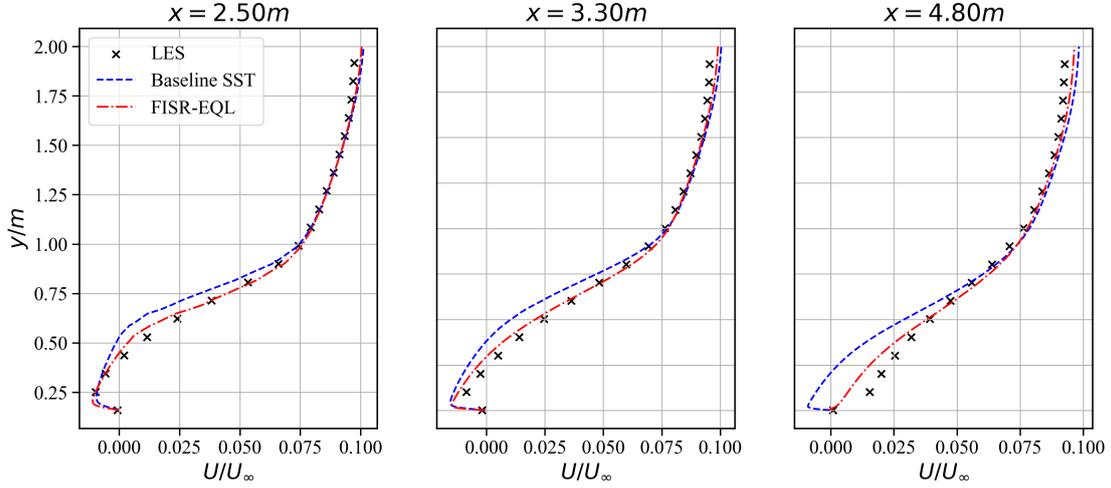

(b) CBFS

Figure 9 Velocity profiles of two training cases

Table 5 reports the relative error of the predicted reattachment location for the different methods, normalized such that the baseline SST–LES discrepancy is set to unity. All approaches mitigate the excessive separation and delayed reattachment predicted by the original SST model. The two-stage FIML-classic strategy achieves substantial improvement during the first inversion step; however, once the neural network is trained offline and redeployed, the predictive accuracy deteriorates noticeably. The explicit expression obtained using the proposed FISR-EQL framework yields a larger error compared with FIML-direct. This outcome is expected, as the neural network in FIML-direct contains substantially more parameters than the sparse structure produced by EQL and therefore has greater capacity to fit the training data. From an optimization perspective, FISR-EQL must additionally balance the sparsity objective, which can lead to a modest degradation in QoI performance. Nevertheless, as discussed earlier, the resulting analytical expression offers clear advantages in terms of interpretability and practical deployability over neural-network-based models.

Table 5 Summary of reattachment point error for all cases

| Method | Trainable Parameter | Hump | CBFS |
|---|---|---|---|
| **Baseline SST** | N/A | 100% | 100% |
| **Field Inversion** | 81157 | 4.29% | 2.14% |
| **FIML-classic** | 1801 | 8.80% | 14.81% |
| **FIML-direct** | 1801 | 1.42% | 10.27% |
| **FISR-EQL** | 9 (Sparse) / 202 (original) | 0.28% | 32.99% |

The spatial distribution of the correction field is illustrated in Figure 10. The learned expression obtained from the training set accurately identifies the separated shear-layer region and applies an appropriate correction magnitude, thereby reducing the excessively large separation bubble.

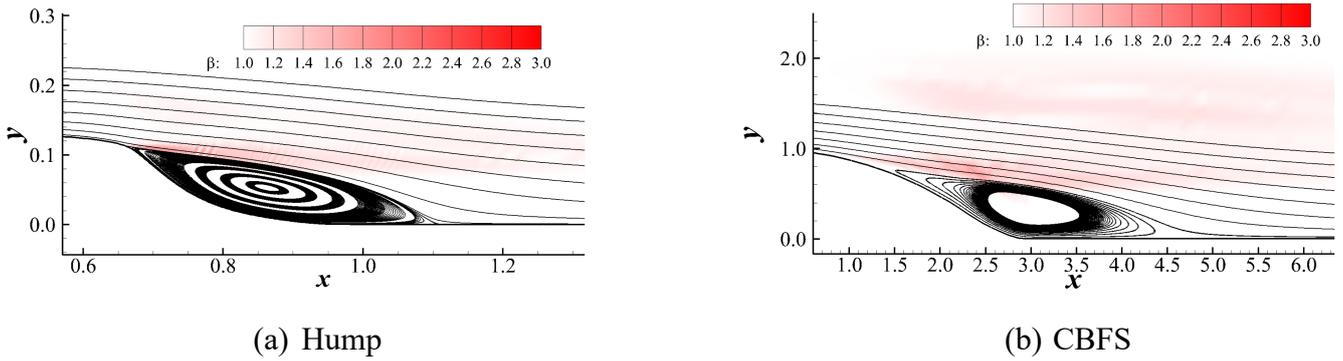

(a) Hump          (b) CBFS

Figure 10 Streamline and $\beta$ distribution of two training cases

## 4. Test Cases

This section evaluates the learned expression obtained in Section 3 to demonstrate its strong generalization capability for separated shear flows. The selected test suite includes the canonical periodic-hill case, a three-dimensional cube flow, and the two-dimensional high-lift NLR7301 airfoil.

### 4.1 Periodic Hill

The periodic-hill configuration and its DNS reference data were generated by Xiao et al.[45] Three geometries with different slope parameters are considered in the present study, as shown in Figure 11.

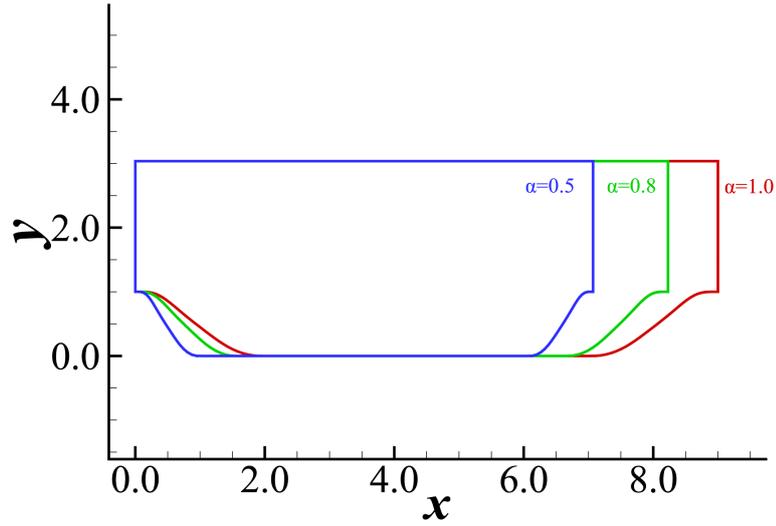

Figure 11 Periodic Hill Geometry

The Reynolds number, based on the hill height, is 5600. The computational mesh contains approximately 14,000 cells, and simulations are performed using simpleFoam[46]. Taking the case with $\alpha = 0.8$ as an example, the resulting effective $\beta$ field is presented in Figure 12.

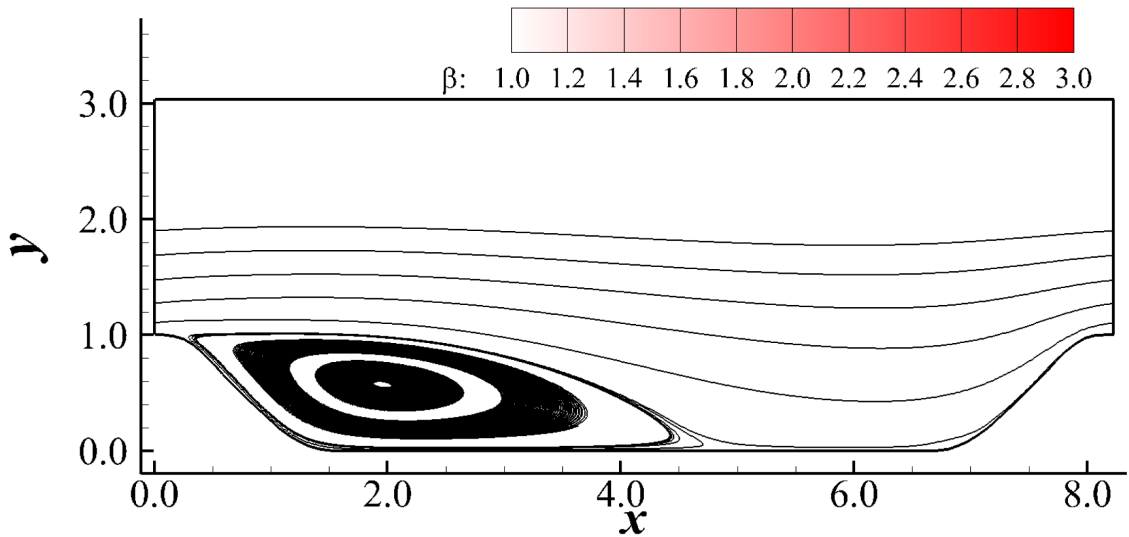

(a) DNS[45]

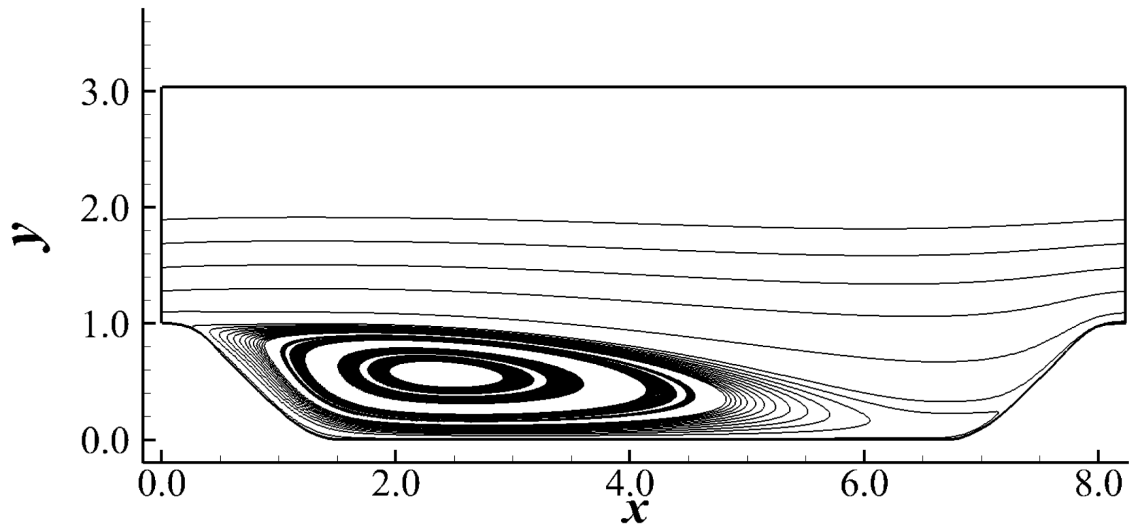

(b) Baseline SST

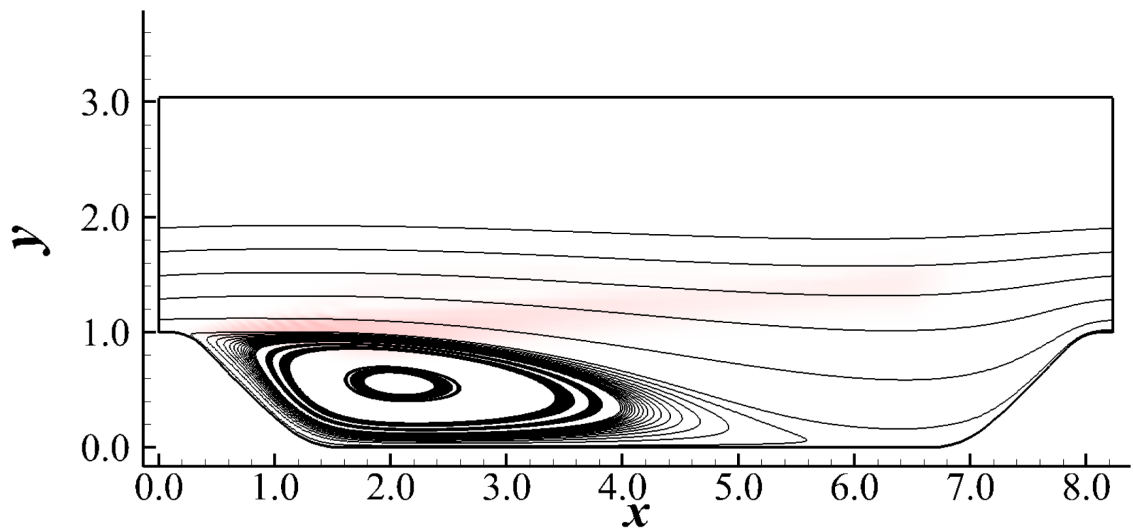

(c) FISR-EQL

Figure 12 streamline of periodic Hill

Consistent with the behavior observed in the training cases, the correction is activated primarily within the free shear layer and significantly alleviates the excessive separation bubble predicted by the baseline SST model. Figure 13 shows the velocity profiles within the flow field, where the proposed model reduces the discrepancy relative to the reference data. The root-mean-square error (RMSE), computed according to the following definition,

$$\text{RMSE} = \sqrt{\frac{1}{N}\sum_{i=1}^{N}\left(U_i^{\text{RANS}} - U_i^{\text{ref}}\right)^2} \quad (15)$$

is reduced by 12.02% ($\alpha = 0.5$), 31.84% ($\alpha = 0.8$), and 29.41% ($\alpha = 1.0$). Here, $U_i^{\text{RANS}}$ and $U_i^{\text{ref}}$ denote the predicted and reference (DNS) streamwise velocities at the i-th sampling location, respectively, and N is the number of sampling points.

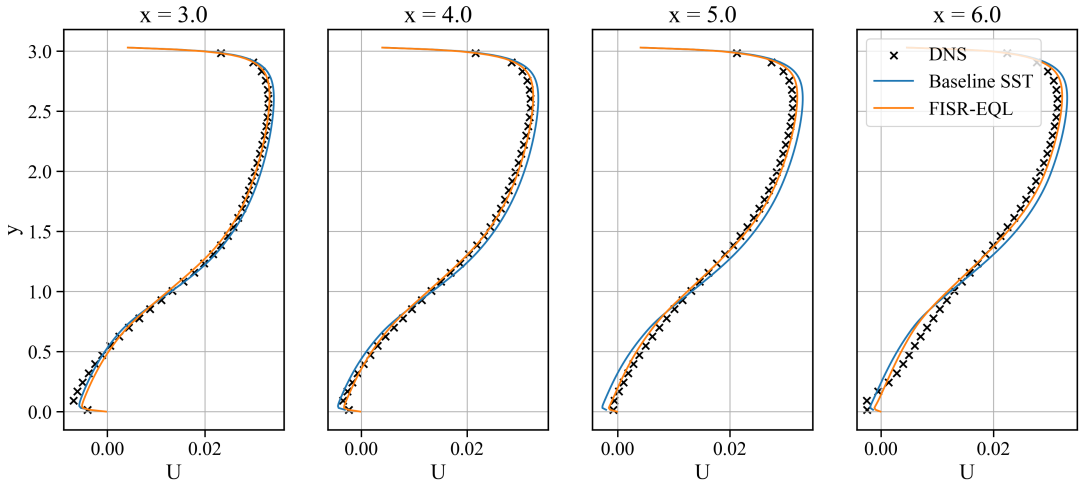

(a) $\alpha = 0.5$

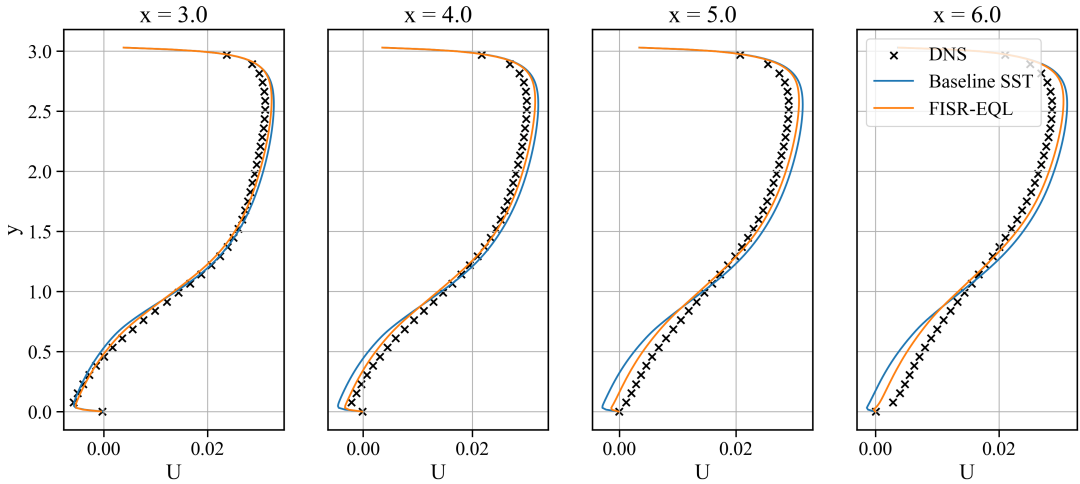

(b) $\alpha = 0.8$

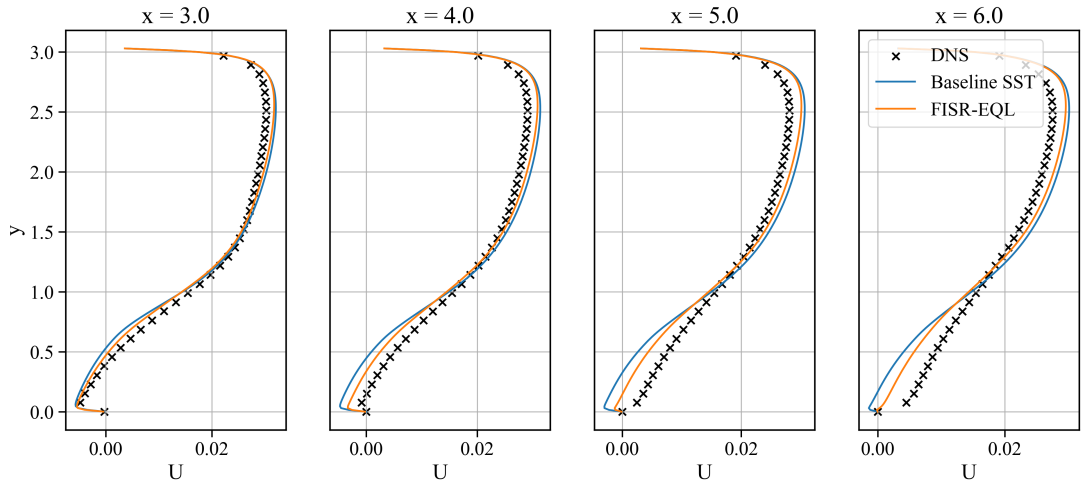

(c) $\alpha = 1.0$

Figure 13 U profile of Periodic Hill

### 4.2 NLR7301 Two-Element Airfoil

For high-lift configurations, the SST model typically overpredicts the extent of flow separation and consequently underestimates the stall angle.[47] The NLR7301 two-element airfoil is a canonical configuration for assessing the predictive accuracy of flap aerodynamics.[48,49] The computational mesh used in the present study is shown in Figure 14. The Reynolds number is $Re = 2.51 \times 10^6$ and the Mach number is $Ma = 0.185$. Because compressibility effects are non-negligible for this case, simulations are performed using HiSA.[50]

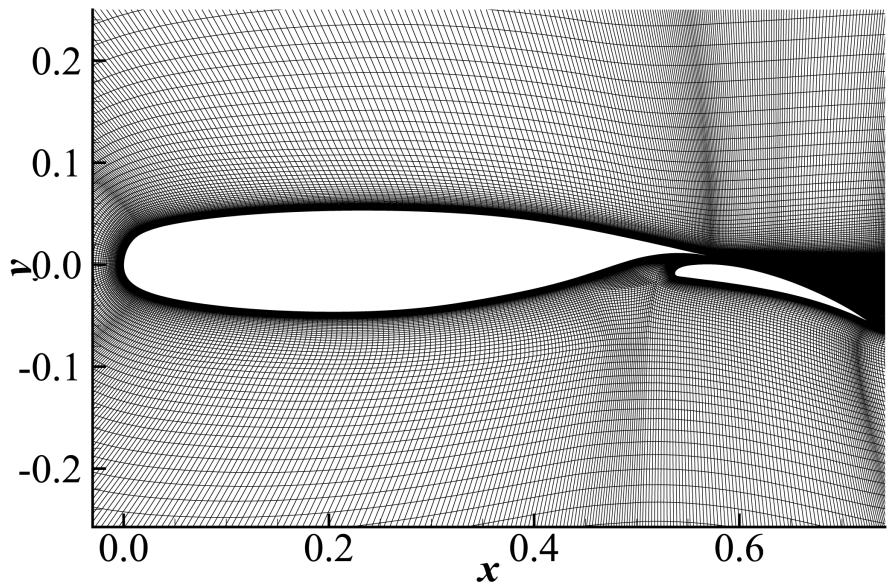

Figure 14 Geometry and Mesh of NLR7301

The lift coefficient as a function of angle of attack is presented in Figure 15. The proposed model accurately predicts both the stall angle and the maximum lift coefficient. In the linear regime, the drag prediction remains consistent with the baseline SST model, indicating that the boundary layer is properly preserved.

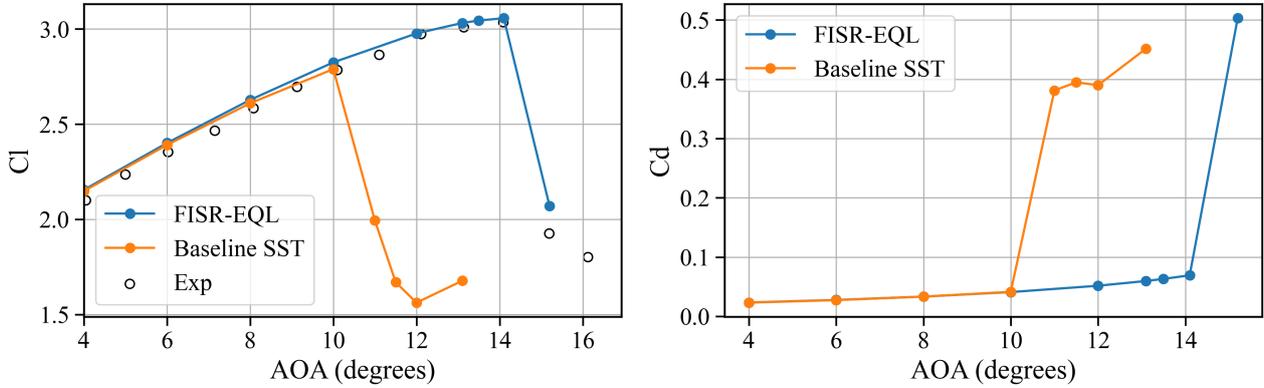

Figure 15 $C_L$ and $C_D$ to Angle of Attack Comparison

The correction near the flap region is illustrated in the contour plots in Figure 16. Because the baseline SST model has already stalled at this angle of attack, unsteady separation occurs near the trailing edge. As a result, the computation in Figure 16(a) did not converge, and only the streamlines from a single time step are shown. The stall of high-lift devices shares certain similarities with geometry-induced separation. Therefore, by increasing the eddy viscosity in regions with strong shear, the FISR-EQL model can delay separation and thus correct the prematurely predicted stall.

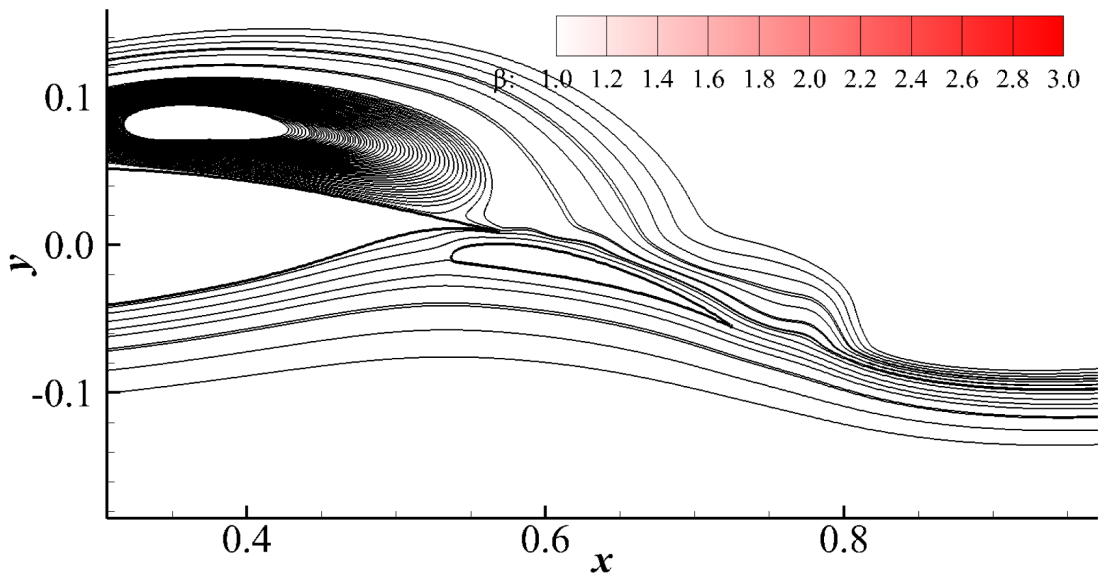

(a) Baseline SST model ($\alpha_{AoA} = 13.5°$)

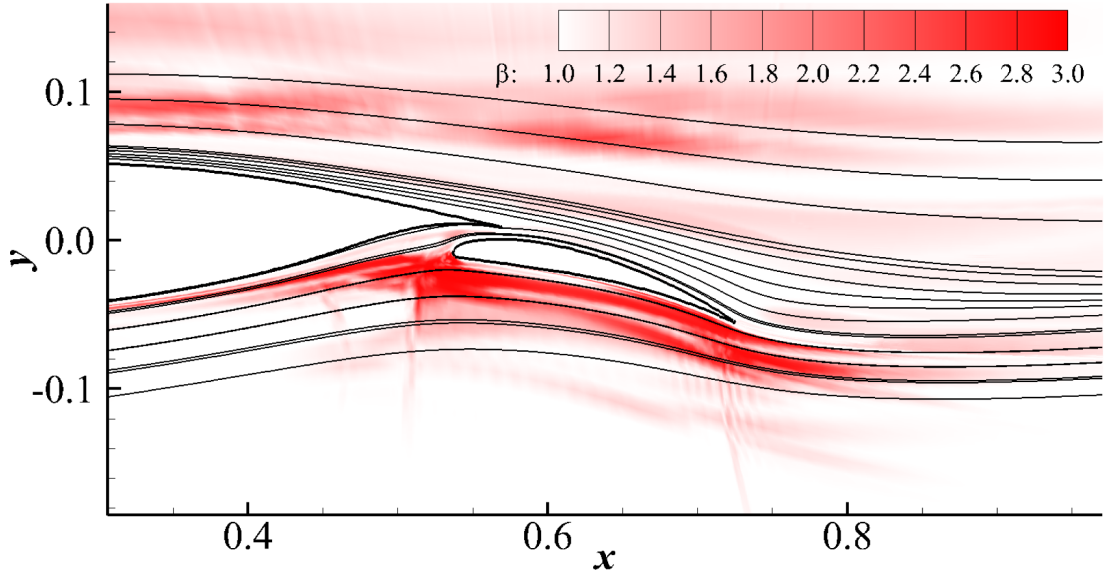

(b) FISR-EQL ($\alpha_{AoA} = 13.5°$)

Figure 16 $\beta$ Contour and Streamline in Flap Region of NLR7301

## 4.3 FAITH Hill

In this section, the proposed model is applied to the FAITH Hill case[51,52], which provides a test for turbulence models due to the presence of three-dimensional separation and complex reattachment behavior. The geometry of this case is shown in Figure 17. The CFD setup is consistent with the experiment. The bump height is $H = 0.1524m$, and the Reynolds number based on $H$ is $5 \times 10^5$. A half-domain configuration is employed, with a symmetry boundary condition applied at the mid-plane. The upper and lower surfaces, as well as the outer boundary opposite to the symmetry plane, are treated as no-slip wall boundaries. A velocity profile is prescribed at the inlet, with a mean velocity of 50.292 m/s, consistent with the experimental conditions.

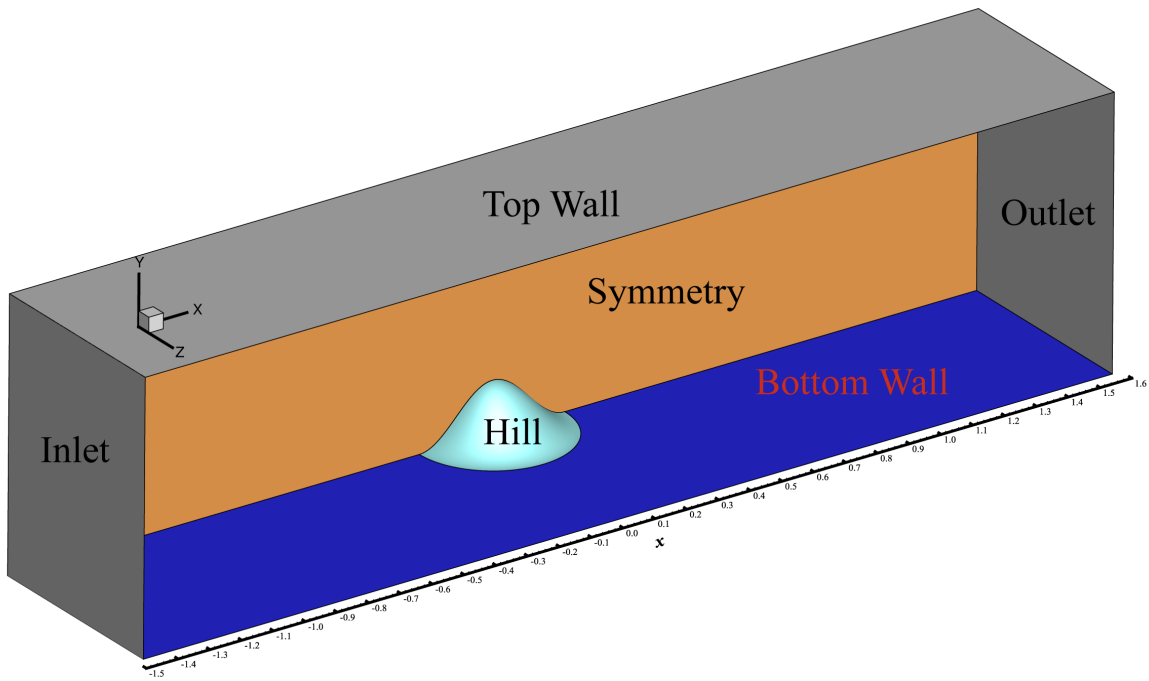

Figure 17 Geometry of FAITH Hill

To assess the capability of the model, detailed comparisons are performed against DNS data and baseline RANS results. In particular, streamwise velocity profiles at several representative streamwise locations are examined to evaluate the model's ability to capture the development of the separated shear layer and the recovery process downstream of reattachment. Figure 18 presents the velocity profile distributions at different streamwise locations on the symmetry plane. The neural-network-based FIML-classic and FIML-direct models are barely activated; in regions where corrections are expected, they output values close to 1, resulting in predictions that are essentially identical to those of the SST model. This is because when the distribution of input features shifts away from that of the training set, neural networks tend to diverge or produce other unpredictable results, whereas explicit expressions exhibit greater robustness. After correction, the FISR-EQL model reduces the mean velocity error at these locations by 29.67% compared to the SST model.

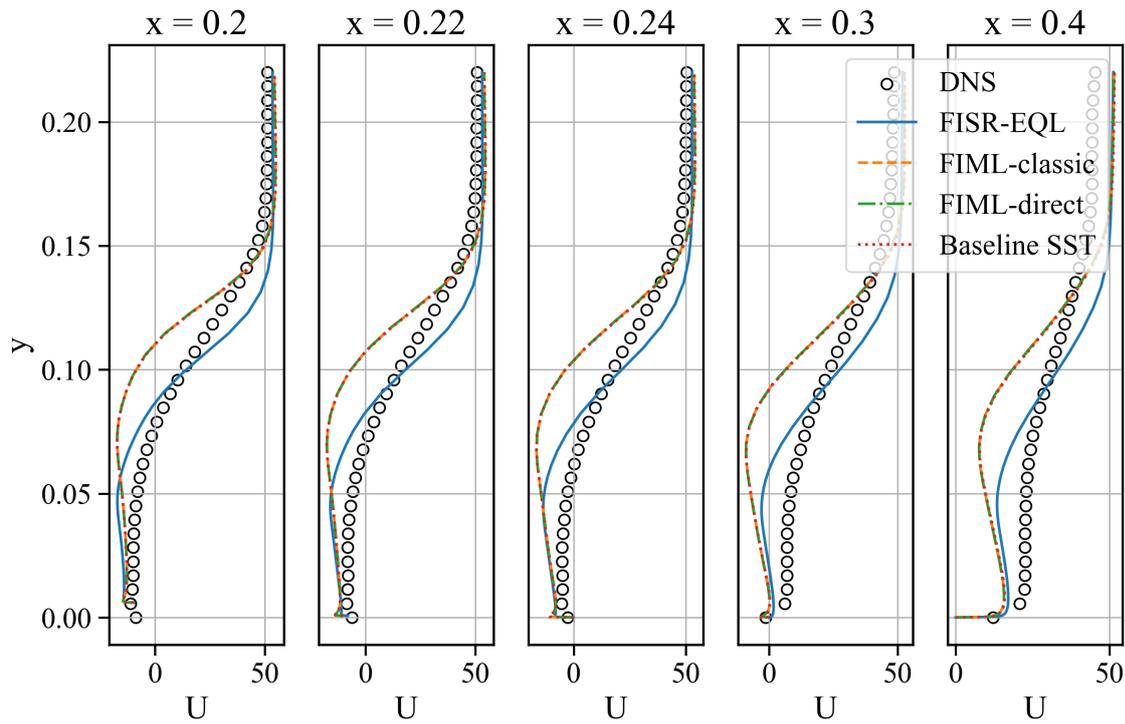

Figure 18 U profile of FAITH Hill

Figure 19 shows the distribution of the correction field on the symmetry plane. The correction remains activated in regions with strong shear to suppress premature separation, thereby reducing the separation region. This suppression effect is evident in Figure 20. Compared with the SST model, the FISR-EQL model alleviates premature separation, making the zero-crossing point of the skin-friction coefficient closer to the experiment result.

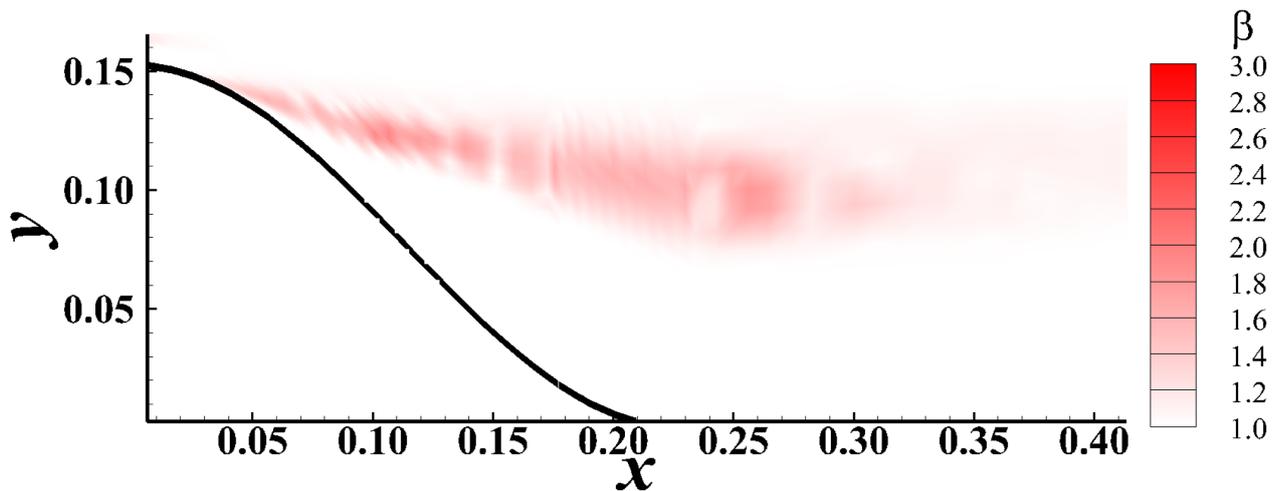

Figure 19 $\beta$ Distribution on symmetry plane of FAITH Hill

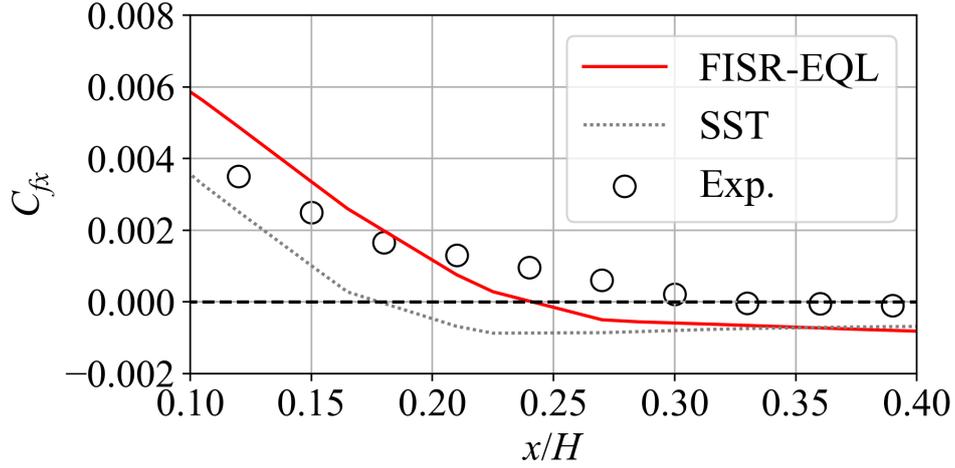

Figure 20 $C_{fx}$ Distribution on Symmetry Plane

**4.4 Zero-Pressure-Gradient (ZPG) Flat-Plate Boundary Layer**

Section 4.3 showed that, for the NLR7301 case at low angles of attack, the FISR-EQL model and the baseline SST model yield essentially identical predictions of the skin-friction drag coefficient, confirming the effectiveness of the shielding function in preserving boundary-layer behavior. In this section, we further assess boundary-layer protection using a ZPG flat-plate boundary-layer benchmark. Skin-friction distributions and velocity profiles are reported to verify that the correction does not contaminate attached boundary layers.

Simulations are performed using simpleFoam. The Reynolds number based on the flat-plate length is $Re_L = 1.0 \times 10^7$, and the mesh consists of approximately $2 \times 10^5$ rectangular cells. Figure 21(a) demonstrates that the FISR-EQL model predicts the skin-friction coefficient distribution with the same accuracy as the baseline SST model, while Figure 21(b) shows that both approaches reproduce the boundary-layer velocity profile with excellent agreement.

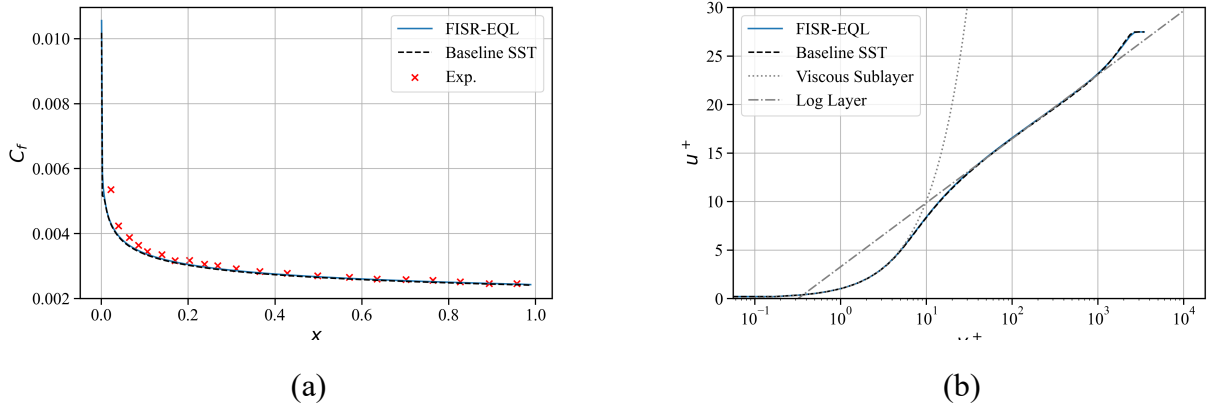

(a)         (b)

Figure 21 $C_f$ distribution and Velocity Profile of boundary layer

# 5. Summary

This work presents FISR-EQL, an end-to-end, interpretable turbulence model correction framework that embeds an EQL directly into a PDE-constrained optimization process based on the adjoint method. Unlike conventional two-stage FIML and FISR approaches, which separate field inversion and surrogate modeling, the proposed method optimizes the correction model directly in parameter space using discrete adjoint gradients, thereby eliminating objective mismatch while maintaining consistency with the governing equations.

Applied to the SST model through a correction to the turbulent kinetic energy production term, the framework incorporates a shielding mechanism to preserve attached boundary-layer behavior. The embedded EQL architecture promotes sparsity during training, enabling the extraction of compact analytical expressions rather than black-box neural networks.

Training on the CBFS and NASA hump cases demonstrated substantial improvement over FIML-classic and performance comparable to FIML-direct, while retaining interpretability. The learned expression generalized well to unseen configurations, including periodic hills, a surface-mounted cube, and the NLR7301 high-lift airfoil, improving separation prediction and stall characteristics without degrading attached-flow drag.

Overall, this study presents a preliminary attempt to apply the FISR-EQL approach to steady-state flow problems and reconciles optimality and interpretability in data-driven turbulence modeling, providing a practical pathway toward transparent and generalizable RANS model corrections. This framework shows significant potential for model correction and optimization in unsteady problems. Future work will further explore its application to various unsteady cases while also investigating more effective feature selection strategies.

# Acknowledgments

This work was supported by the National Natural Science Foundation of China (grant numbers 12372288, U2541235, U23A2069 and 12388101); and the National Key Research and Development Program of China (2024YFB4205601) and other national research projects.